\documentclass[twocolumn,aps,prl,floatfix,nofootinbib]{revtex4-2}
\usepackage{amsmath}
\usepackage{amsfonts}
\usepackage{amssymb}
\usepackage{graphicx}
\usepackage{color}
\usepackage{bm}

\usepackage{soul}

\usepackage{mathtools}
\DeclarePairedDelimiter\bra{\langle}{\rvert}
\DeclarePairedDelimiter\ket{\lvert}{\rangle}
\DeclarePairedDelimiterX\braket[2]{\langle}{\rangle}{#1 \delimsize\vert #2}
\usepackage[final]{hyperref} 
\hypersetup{
	colorlinks=true,       
	linkcolor=blue,        
	citecolor=blue,        
	filecolor=magenta,     
	urlcolor=blue         
}

\begin{document}

\title{Mpemba effects in open nonequilibrium quantum systems}

\author{Andrea Nava}
\author{Reinhold Egger}
\affiliation{Institut f\"ur Theoretische Physik, Heinrich-Heine-Universit\"at, D-40225  D\"usseldorf, Germany}

\begin{abstract}
We generalize the classical thermal Mpemba effect (where an initially hot system relaxes faster to the final 
equilibrium state than a cold one) to open quantum systems coupled to several reservoirs.  We show that, in general, two different types of quantum Mpemba effects are possible.  They may be distinguished by quantum state tomography.  However, the existence of a 
quantum Mpemba effect (without determining the type) can already be established by measuring simpler observables such as currents
or energies.  We illustrate our general results for the experimentally feasible case of 
an interacting two-site Kitaev model coupled to two metallic leads.  
\end{abstract}
\maketitle

\emph{Introduction.---}Quantum versions of the classical thermal Mpemba effect (ME) \cite{Mpemba_1969,Lu2017}
have attracted considerable recent attention, see, e.g., Refs.~\cite{Murciano_2024,Rylands2023,Shapira2024,Joshi2024,Chatterjee2023,Chatterjee2023_2,Wang2024,Strachan2024,liu2024,Turkeshi2024,Kochsiek2022,Klich2019,Ares2024}.  
The classical ME, investigated and observed in a wide variety of systems \cite{Lasanta2017,Jesi2019,Torrente2019,Santos2020,Megias2022,Chetrite2021,Walker2023}, arises when two copies of the 
same system are prepared in an equilibrium state at temperatures $T_h$ (``hot'') and $T_c$ (``cold''),
respectively. For each copy, a sudden quench to the temperature $T_{\rm eq}<T_c<T_h$ is then performed.  Measuring the relaxation times $\tau(T_{h/c})$ towards the final equilibrium configuration with temperature $T_{\rm eq}$, the ME takes place if the corresponding pathways in the energy landscape \cite{Ibanez2024,Adalid2024} are such that the hot
system relaxes faster than the cold one, i.e., for $\tau(T_h)<\tau(T_c)$.
The inverse ME is defined by $\tau(T_c)<\tau(T_h)$ for $T_c<T_h<T_{\rm eq}$.
Conventionally, $\tau$ is determined by the respective time to undergo a phase transition, e.g., between water and ice \cite{Mpemba_1969}, or paramagnetic and ferromagnetic phases \cite{Nava2019,Vadakkayil2021}. For systems without phase transition, the time to reach the final equilibrium state must be
extracted from a suitable ``monitoring function'' \cite{Lu2017}.

Different variants of quantum Mpemba effects (QMEs) have been proposed and studied during the past few years.  
The case of \emph{closed} quantum systems has been addressed, e.g., in Refs.~\cite{Murciano_2024,Rylands2023,Turkeshi2024},
where experimental observations are already available for trapped ions \cite{Shapira2024,Joshi2024}. 
We here instead investigate the QME for \emph{open nonequilibrium} quantum systems coupled to (at least) two different reservoirs (``baths''), where the competition of stochastic relaxation processes and quantum effects   
 drives the system towards a (nonequilibrium or equilibrium) stationary state, dubbed
(N)ESS in what follows.  In this Letter, we introduce a general protocol to unambiguously identify the QME in open nonequilibrium systems with Markovian dynamics. Recent theoretical work \cite{Chatterjee2023,Chatterjee2023_2,Wang2024} proposed to search for (single or multiple) intersection points between time-dependent averages of some system observable computed at different temperatures. This definition is misleading and in conflict with previous work for the classical ME \cite{Holtzman2022,Lu2017}. In fact, for a poorly chosen monitoring function, it causes false QME identification and/or it
misses cases where the QME actually occurs; for details, see the Supplementary Material (SM) \cite{SM}. 
Related work may violate the positivity of the density operator \cite{Strachan2024}.  

We here (i) formulate a general protocol for identifying and classifying the QME in open quantum systems connected to several baths (labeled by index $\lambda$), and (ii) illustrate this protocol for a relatively simple model which can be experimentally realized. Possible extensions to the non-Markovian case \cite{breuer2009, luo2023} are discussed in the SM \cite{SM}.
In general, we consider a set of pre-quench bath parameters $\left\{ p_i \right\}$.  At time $t=0$, one performs a quench
to the after-quench parameters $\{ p_{i,{\rm (N)ESS}} \}$ describing the final (N)ESS configuration.  
The above parameter sets may include, e.g., the chemical potentials $\mu_\lambda$ and thermal energies $k_B T_\lambda$ of each bath. We next extend the notion of hot vs cold initial configurations \cite{Mpemba_1969} to ``far'' vs ``close'' initial conditions ($\{p_{i,f}\}$ vs $\{ p_{i,c}\}$), to describe more general cases where several control parameters are changed. 
For fixed after-quench parameters, we use a Euclidean distance in this parameter space,
\begin{equation}\label{Eucldist}
\mathcal{D}_E \left(\left\{ p_i \right\}\right)=\sqrt{\sum_i \left(p_i-p_{i,{\rm (N)ESS}}\right)^2}.
\end{equation} 
For $\mathcal{D}_E (\left\{ p_{i,c} \right\})<\mathcal{D}_E (\left\{ p_{i,f} \right\})$,
the set $\{p_{i,c}\}$ is considered to be closer to $\{ p_{i,{\rm (N)ESS}}\}$ than the set $\{p_{i,f}\}$.  
Below we also employ the trace distance \cite{Nielsen2000},
\begin{equation}\label{tracedist}
\mathcal{D}_T (\rho (t))=\frac12\mathrm{Tr}\left|\rho(t)-\rho_{\rm (N)ESS}\right|,
\end{equation}
which measures the distance between the system density matrix $\rho(t)$ and the final (N)ESS state $\rho_{\rm (N)ESS}$. 

The choice of the distance functions in parameter [Eq.~\eqref{Eucldist}]  and state [Eq.~\eqref{tracedist}] space is not unique. However, our definition of the QME as given below is robust to changes of the distance functions if reasonable physical requests are satisfied, as is also the case for the original Markovian ME \cite{Lu2017}. In particular, ${\cal D}_E(\{ p_i\})$ must preserve the order of each pre- and post-quench parameter, where the Euclidean distance \eqref{Eucldist}
is a convenient measure.  Furthermore, ${\cal D}_T(\rho(t))$ must be a monotonically non-increasing, continuous, and convex function of time, cf.~Ref.~\cite{Lu2017}. We have chosen the trace distance in Eq.~\eqref{tracedist}, which satisfies all consistency relations under Markovian dynamics \cite{Nielsen2000,Wang2009}, for its operational relevance to many quantum information protocols \cite{Coles2019}. 
A detailed discussion of distance measures \cite{Carollo_2021,Tejero2024,moroder2024, ruskai2022,hermes2017,spehner2013,petz1996,tamir2015} is given in the SM \cite{SM}.

\begin{figure}
\begin{center}
    \includegraphics[width=0.45\textwidth]{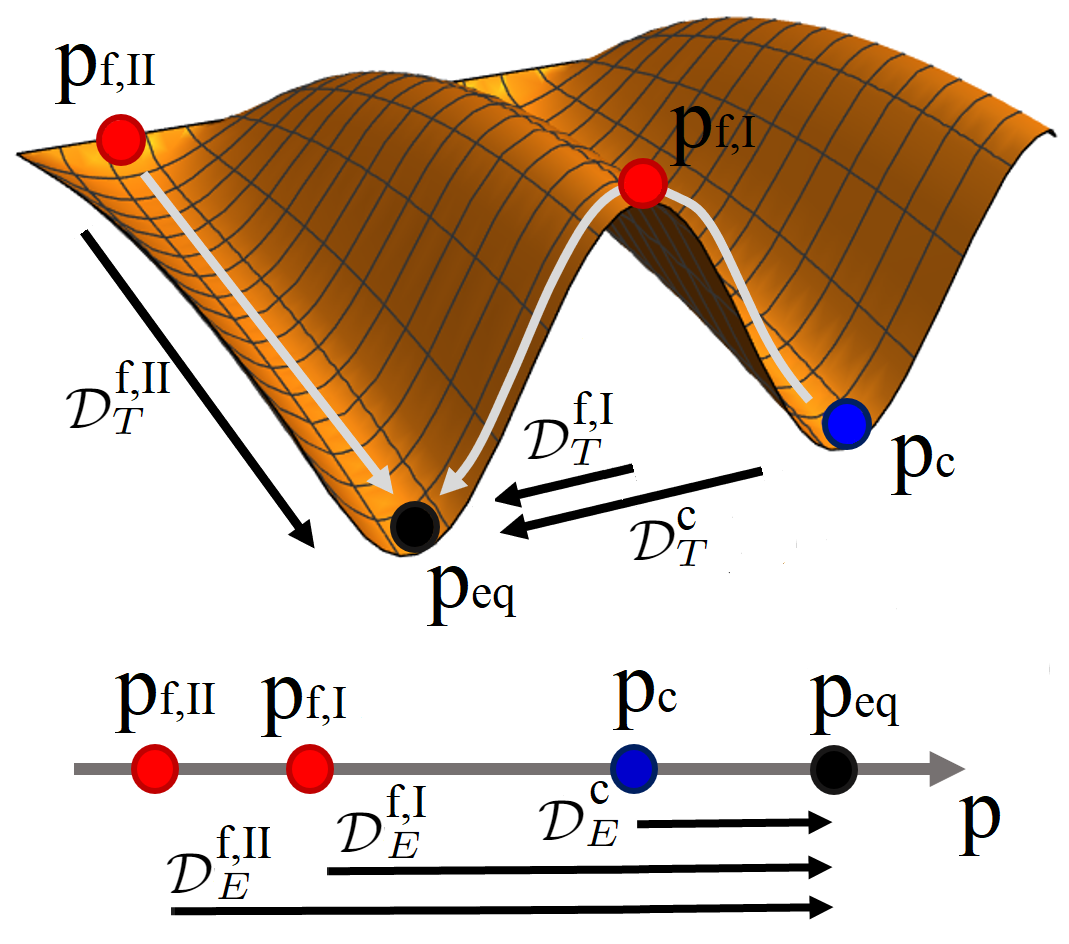}
    \caption{Upper panel: Sketch of the energy landscape in state space and corresponding time-evolution trajectories for type-I and type-II QMEs. The trace distance \eqref{tracedist} between different 
    initial states and the final (N)ESS density matrix (corresponding to $\{p_{\rm eq}\}$) is indicated. 
    We consider ``far'' ($\{p_{i,f}\}$) and ``close'' ($\{p_{i,c}\})$ initial configurations, cf. the lower panel.
    For a type-I QME, the trace distance \eqref{tracedist} satisfies $\mathcal{D}_T (\rho_f (t)) < \mathcal{D}_T (\rho_c (t))$ at all times.  For the more elusive type-II QME, we have $\mathcal{D}_T (\rho_f (t))>\mathcal{D}_T (\rho_c (t))$ at short times $t<t^*$, but the inequality is reversed at longer times.
    Lower panel: In both type-I and type-II cases,  $\mathcal{D}_E (\left\{ p_{i,f} \right\}) > \mathcal{D}_E (\left\{ p_{i,c} \right\})$, see Eq.~\eqref{Eucldist}, which defines far vs close initial configurations.  }
    \label{fig1}
\end{center}
\end{figure}

Two definitions for the classical ME have been introduced in Refs.~\cite{Lu2017,Holtzman2022,Degunther2022} which we unify and extend to the quantum case, see Fig.~\ref{fig1}.  
We label them type-I and type-II QME.
In both cases, the two initial configurations (far vs close) satisfy
$\mathcal{D}_E (\left\{ p_{i,f} \right\}) > \mathcal{D}_E (\left\{ p_{i,c} \right\})$, see Eq.~\eqref{Eucldist}.
For the type-I QME, the initially far system $\{p_{i,f}\}$ reaches the final (N)ESS faster \textit{because}
the corresponding state $\rho_f(0)$ is actually closer to $\rho_{\rm (N)ESS}$,
i.e., $\mathcal{D}_T (\rho_f (0))<\mathcal{D}_T (\rho_c (0))$, see Eq.~\eqref{tracedist}. It thus enjoys a natural advantage over the
nominally closer system during the ensuing time evolution \cite{Holtzman2022}.
For the more elusive type-II QME, the far vs close systems evolve through different paths but \textit{even though} $\rho_c(0)$ is nearer to the (N)ESS state, i.e., $\mathcal{D}_T (\rho_f (0))>\mathcal{D}_T (\rho_c (0))$, the path that connects the far system to the (N)ESS is eventually shorter \cite{Lu2017}.
By analyzing both $\mathcal{D}_T (\rho (t))$ as a function of time as well as $\mathcal{D}_E (\left\{ p_{i} \right\})$, both types of QME can be unambiguously identified.

 Our protocol avoids any false QME detection or mislead since the trace distance \eqref{tracedist}, which is in the range $\left[ 0,1 \right]$, is a monotonically decreasing function of time under Lindblad dynamics \cite{Wang2009}. In particular, (a) if $\mathcal{D}_T (\rho_f (t))>\mathcal{D}_T (\rho_c (t))$ for all times, no QME takes place; (b) if $\mathcal{D}_T (\rho_f (t))<\mathcal{D}_T (\rho_c (t))$ for all times, we have a type-I QME; and (c) if $\mathcal{D}_T (\rho_f (0))>\mathcal{D}_T (\rho_c (0))$ but a finite time $t^*$ exists such that $\mathcal{D}_T (\rho_f (t))<\mathcal{D}_T (\rho_c (t))$ for $t>t^*$, a
type-II QME takes place.  
Below we illustrate both types of QME for a simple and experimentally accessible model, outlining concrete experimental protocols.

\emph{Model.---}We consider the minimal interacting two-site Kitaev model (I2KM) \cite{Leijnse2012,Tsintzis2024} with a non-local Coulomb interaction strength $U$ \cite{Samuelson2024} (putting $\hbar=k_B=1$ below),
\begin{eqnarray} \label{two-site}
H_{\rm{I2KM}}&=& \epsilon_1 n_1 + \epsilon_2n_2 +U n_1 n_2  \\  \nonumber
&+&t_h (c_1^{\dagger} c_2 +c^\dagger_2 c_1^{})+ \Delta c_1^{\dagger}c_2^{\dagger}+ \Delta^* c_2^{} c_1^{},
\end{eqnarray}
where $c_i^{}$ and $n_i=c_i^{\dagger} c_i^{}$ are spinless electron annihilation and occupation number operators for site (quantum dot) $i\in\{1,2\}$, respectively.  
The $\epsilon_i$ are on-site energies, $t_h$ the tunneling amplitude connecting the two dots, $\Delta$ a superconducting pairing amplitude due to crossed Andreev reflection processes. 
Both $t_h$ and $\Delta$ can be mediated by a short superconductor, $\epsilon_i$ may be controlled by voltages applied to finger gates.  The I2KM has recently 
been studied experimentally in the context of realizing ``poor man's'' Majorana bound states \cite{Dvir2023,Bordin_2024,tenHaaf2024}. We therefore expect that our predictions on the QME can be readily put to a test. Using the basis 
\begin{equation}\label{many-body-basis}
    \{ \ket{0}, \,\ket{1}=c_1^{\dagger}\ket{0}, \,\ket{2}=c_2^{\dagger}\ket{0}, \,
\ket{d}=c_1^{\dagger}c_2^{\dagger}\ket{0}\},
\end{equation} 
Eq.~\eqref{two-site} is expressed as
\begin{equation} \label{many-body-two-site}
H_{\rm I2KM}=\left(\begin{array}{cccc}
0 & 0 & 0 & \Delta^{*}\\
0 & \epsilon_{1} & t_h & 0\\
0 & t_h & \epsilon_{2} & 0\\
\Delta & 0 & 0 & \epsilon_{d}
\end{array}\right),
\end{equation}
with $\epsilon_{d}=\epsilon_{1}+\epsilon_{2}+U$. The states $|0\rangle$ and $|d\rangle$ ($|1\rangle$ and $|2\rangle$) have even (odd) fermion parity.

\begin{figure}
\begin{center}
    \includegraphics[width=0.4\textwidth]{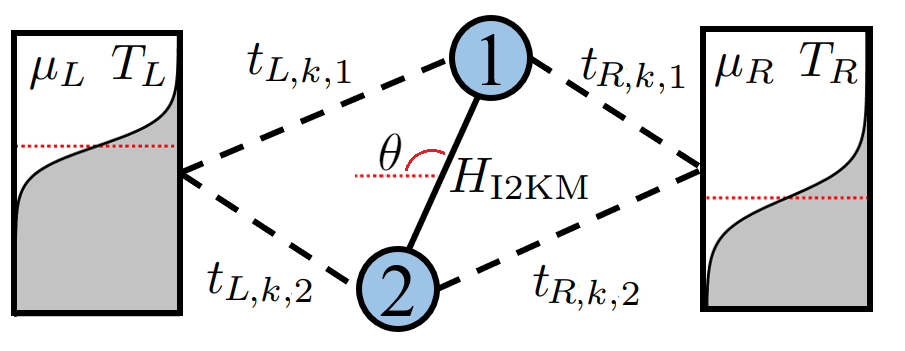}
    \caption{Two quantum dots described by $H_{\rm I2KM}$ are coupled to left and right 
    ($\lambda=L,R$) metallic leads (Fermi gases) with the respective
    temperature $T_{\lambda}$ and chemical potential $\mu_{\lambda}$.  Fermi distributions in both leads are schematically indicated, with $\mu_\lambda$ shown as red dotted lines. Electrons  with momentum $k$ in lead $\lambda$ are tunnel-coupled to dot $i\in \{1,2\}$ with amplitude $t_{\lambda,k,i}$. The angle $\theta$ defines the relative position between the dots and the leads.}
    \label{fig2}
\end{center}
\end{figure}

Next, see Fig.~\ref{fig2}, the I2KM is coupled to left and right $(\lambda=L,R)$ metallic leads by tunnel contacts ($H_c)$.  Describing the leads by noninteracting fermions ($H_l$) with chemical potential $\mu_\lambda$ and temperature $T_\lambda$, the total Hamiltonian is 
$H=H_{\rm{I2KM}}+H_{l}+H_{c}$ with $H_{l}=\sum_{\lambda,k} \epsilon_k^{} c_{\lambda,k}^{\dagger} c^{}_{\lambda,k}$ and 
$H_{c}=\sum_{\lambda,k,i} (t_{\lambda,k,i}^{} c_{\lambda,k}^{\dagger} c_{i}^{} +\text{h.c.})$.
Here $c_{\lambda,k}$ are electron annihilation operators for lead $\lambda$ and momentum $k$ (with dispersion $\epsilon_k)$,
and $t_{\lambda,k,i}$ is a tunneling amplitude connecting the respective lead electron to dot $i$.  
The angle $\theta$ encodes the relative orientation of the I2KM with respect to the leads, see Fig.~\ref{fig2}. 

\emph{Lindblad equation (LE).---}For weak tunnel couplings $t_{\lambda,k,i}$, 
a LE is expected to describe the time evolution of the I2KM density matrix $\rho(t)$  \cite{Benenti2009,Bauer2011,Nava2021,Artiaco2023}. Integrating out the leads in the wide-band approximation \cite{Nazarov2009,Nakajima2015,Cao2017}, under standard Born-Markov and rotating-wave approximations \cite{breuer2007theory}, we obtain a LE,
\begin{equation} \label{lindblad}  
    \partial_t \rho = -i\left[H_{\mathrm{I2KM}},\rho\right]+  \sum_{m\neq n} 
    \Gamma_{m,n} \,{\cal L}\left[L_{m,n}\right]\rho,
\end{equation}
with the dissipator ${\cal L}[L]\rho= L\rho L^\dagger - \frac12 \{ L^\dagger L,\rho\}$.
For a given jump operator $L_{m,n}=\ket{m}\bra{n}$, with $m$ and $n$ running over the I2KM many-body states  
 in Eq.~\eqref{many-body-basis}, $\Gamma_{m,n}=\sum_{\lambda=L,R}\Gamma_{m,n}^\lambda$ denotes the corresponding transition rate. 
Within the Lindblad approach, the leads act locally \cite{Tupkary2023,Oliveira2016,Cattaneo2019} injecting (or removing) a single electron into (from) the I2KM at a time.
The non-vanishing transition rates in Eq.~\eqref{lindblad} only connect states with different parity,  $\{\ket{0},\ket{d}\} \leftrightarrow \{\ket{1},\ket{2}\}$. Conversely, the Hamiltonian $H_\mathrm{I2KM}$ only connects states with same parity, $\ket{1}\leftrightarrow\ket{2}$ or 
$\ket{0}\leftrightarrow\ket{d}$, see Eq.~\eqref{many-body-two-site}.
We find the non-vanishing rates ($i=1,2$)
\begin{equation}\label{ratedef}
\Gamma^\lambda_{i,0}=\gamma_{\lambda,i}(\epsilon_i) f_\lambda(\epsilon_i),\quad  
\Gamma_{d,i}^\lambda =\gamma_{\lambda,3-i}(\epsilon_d-\epsilon_{i}) f_\lambda(\epsilon_d-\epsilon_{i}),  
\end{equation}
with detailed balance relations 
\begin{equation}\label{detbal}
\Gamma^\lambda_{i,0}/\Gamma^\lambda_{0,i}=e^{-(\epsilon_i-\mu_\lambda)/T_\lambda},\quad
\Gamma^\lambda_{d,i}/\Gamma^\lambda_{i,d}=e^{-(\epsilon_d-\epsilon_i-\mu_\lambda)/T_\lambda},
\end{equation}
where $f_\lambda(\epsilon)=[e^{(\epsilon-\mu_\lambda)/T_\lambda} +1]^{-1}$, $\gamma_{\lambda,i}(\epsilon)=2\pi\nu_{\lambda}(\epsilon)\,|t_{\lambda,i}(\epsilon)|^2$, $t_{\lambda,i}(\epsilon_k)=t_{\lambda,k,i}$ and $\nu_{\lambda} (\epsilon)$ the lead density of states.  Following standard arguments \cite{Nazarov2009}, the rates $\gamma_{\lambda,i} (\epsilon)=\gamma_{\lambda,i}$ 
 in Eq.~\eqref{ratedef} are assumed energy-independent. For simplicity, we  parameterize them as 
\begin{equation}
    \gamma_{L,1}=\gamma_{R,2}= \frac{\Gamma}{2} (1+\cos\theta), \quad
    \gamma_{L,2}=\gamma_{R,1}= \frac{\Gamma}{2} (1-\cos\theta).
\end{equation}
For each lead $\lambda$,
the total hybridization strength is thus assumed independent of $\theta$ and $\lambda$, i.e., $\gamma_{\lambda,1}+\gamma_{\lambda,2}=\Gamma$.
Note that our definition of the QME is not limited to a specific model. Here, we focus on a linear (lead-I2KM-lead) geometry, where a LE with local baths is needed, see Eq.~\eqref{lindblad}.
In the SM \cite{SM}, we discuss another model where a global bath applies.
 
\emph{Dynamics.---}The presence of two independent electron reservoirs renders the system evolution 
richer than in the standard single-bath case. If the two baths have the same chemical potential and temperature, 
the I2KM is driven towards an ESS for $t\to\infty$, with zero net current flowing between the leads.
Contrarily, for $\left\{ \mu_L,T_L \right\}\neq \left\{\mu_R,T_R\right\}$, the system evolves 
towards a current-carrying NESS.  For the time-dependent I2KM state $\rho(t)$,
the net current from lead $\lambda$ to the I2KM is given by $I^{\rm tot}_\lambda=I^{\rm in}_\lambda-I^{\rm out}_\lambda$, where
\begin{eqnarray} \nonumber
    I^{\rm in}_\lambda&=&\sum_{i=1,2} \left[\gamma_{\lambda,i} f_\lambda(\epsilon_i)\rho_{0,0}+\gamma_{\lambda,3-i} f_\lambda(\epsilon_d-\epsilon_i)\rho_{i,i} \right], \\ \label{current}
    I^{\rm out}_\lambda&=&\sum_i \left[ \gamma_{\lambda,i} \Bar{f}_\lambda(\epsilon_i)\rho_{i,i}+\gamma_{\lambda,3-i} \Bar{f}_\lambda(\epsilon_d-\epsilon_i)\rho_{d,d} \right]
\end{eqnarray}
with $\Bar{f}_\lambda(\epsilon)=1-f_\lambda(\epsilon)$ and $\rho_{m,n}$ the density matrix elements in the many-body state basis in Eq.~\eqref{many-body-basis}.
The current $I_{1,2}$ between the two dots and the I2KM energy $E$ are 
\begin{equation}\label{current2}
  I_{1,2}=-it_h\left(\rho_{2,1} - \rho_{1,2} \right),\quad   E= \mathrm{Tr}\left(\rho H_{\rm I2KM}\right).
\end{equation}
All the above quantities can monitor the time evolution of the I2KM.

Equation~\eqref{lindblad} can be expressed in the equivalent  superoperator representation, $\partial_t \hat{\rho} = {\bf M} \hat{\rho}$, with $\hat{\rho}$ the $16$-component vectorized form of the $4\times 4$ density matrix and ${\bf M}$ a $16\times 16$ matrix \cite{breuer2007theory}. For given initial state ${\hat{\rho}}(t=0)$, this differential equation system is solved by ($\alpha=1,\ldots,16$)
 \begin{equation}\label{gensol}
     {\hat \rho}_\alpha (t)= \sum_{\beta,\eta=1}^{16} e^{\lambda_\beta t} R_{\alpha,\beta} L_{\beta,\eta}\, \hat{\rho}_\eta (0),
 \end{equation}
with the eigenvalues $\lambda_\beta$ and the corresponding right and left eigenvectors $\mathbf{R}_\beta$ and $\mathbf{L}_\beta$ of ${\bf M}$.  
The time evolution of the system sensitively depends on the eigenvalues $\lambda_\beta$.
The real parts of all $\lambda_\beta$ are non-positive (the LE describes a relaxation dynamics), and complex eigenvalues form complex conjugate pairs (arising from the system Hamiltonian). 
Eigenvalues with ${\rm Re}(\lambda_{\beta})<0$ describe exponential decay, while 
${\rm Im}(\lambda_{\beta})\ne 0$ results in damped (spiral path) or undamped (elliptic path) oscillations \cite{Woolf2009,Lange2021}.At least one  eigenvalue $\lambda=0$ exists. This eigenvalue corresponds to the (N)ESS, 
$\hat{\rho}_{\rm (N)ESS}=\hat{\rho}(t\to\infty)$ with $\partial_t{\hat{\rho}}_{\rm (N)ESS} = 0$. 
If all states are accessible to the Lindblad dynamics in Eq.~\eqref{lindblad}, 
either because of the dissipative transition rates $\Gamma_{m,n}$ or 
because of the unitary dynamics due to $H_{\rm I2KM}$, the stationary state can be reached. In principle, several eigenvalues may vanish, causing a (N)ESS manifold, where the final state depends on the initial condition \cite{Fernengel2020,Rose2016}. 
If the $\Gamma_{m,n}$ can mix all system states, a closed set of 
equations (not depending on the Hamiltonian) holds for the diagonal 
elements of $\rho(t)$ \cite{SM},
giving purely exponential decay. If the dissipative rates cannot trigger all possible 
system transitions, as is the case in our system, 
a competition between coherent Hamiltonian evolution and incoherent dissipative 
evolution takes place. Then the coupled equations for the full density matrix must be solved.

\begin{figure}
\begin{center}
\includegraphics[width=0.44\textwidth]{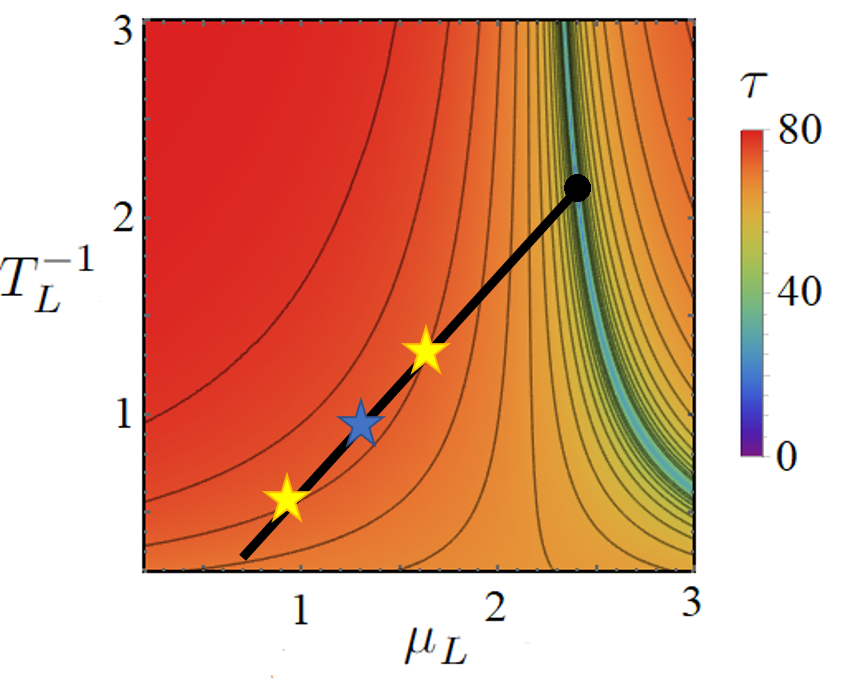}
\end{center}
\caption{Color-scale plot of the I2KM relaxation time $\tau$ towards the ESS (black circle) 
in the $T_L^{-1}$-$\mu_L$ plane for different initial NESS configurations $(T^{-1}_L,\mu_L)$, where $(T^{-1}_R,\mu_R)$ is identical to the final ESS parameters, 
$T^{-1}_{\lambda,{\rm ESS}}=2.15$ and $\mu_{\lambda,{\rm ESS}}=2.4$ for $\lambda=L,R$.
Other parameters are $t_h=1$, $\epsilon_1=\epsilon_2=2$, $\Delta=0.1$, $U=0.25$,  
$\Gamma=0.1$, and $\theta=0$.
Thin black curves indicate $\tau$-isolines.  Yellow and blue stars (located on a thick black line 
connected with the ESS) denote initial states discussed in the main text. 
}\label{fig3}
\end{figure}

\emph{Observing the QME.---}We now discuss typical results for the above model, which were
obtained by numerically solving Eq.~\eqref{lindblad}. Additional results can be found in the SM \cite{SM}. We focus on a specific I2KM parameter set and show the relaxation time $\tau$ from different initial states towards a fixed final ESS configuration in Fig.~\ref{fig3}. Here $\tau$ was
obtained from the time dependence of the trace distance \eqref{tracedist}.  
Figure \ref{fig3} shows a color-scale plot of $\tau$ in the $T_L^{-1}$-$\mu_L$ plane,
where initial NESS configurations are defined by $(T^{-1}_{L},\mu_{L})$, with $(T^{-1}_R,\mu_R)$ identical as for the ESS.  Let us first focus on the two initial states marked by yellow stars in Fig.~\ref{fig3}. Both states are located on a $\tau$-isoline and thus have the same relaxation time even though 
the state at $(T_{L,f}^{-1},\mu_{L,f})=(0.5,0.9)$ on the lower left side is further away 
from the ESS, as measured by ${\cal D}_E$ in Eq.~\eqref{Eucldist}.  
We compare this state to the initial state $(T_{L,c}^{-1},\mu_{L,c})$ marked by a blue star in Fig.~\ref{fig3} which has a longer relaxation time $\tau$ even though it clearly is closer to the ESS.
We thus conclude that the QME takes place.  Comparing instead the other initial
state marked by a yellow star to the one marked by the blue star, no QME occurs. 

\begin{figure}
\begin{center}
\includegraphics[width=0.45\textwidth]{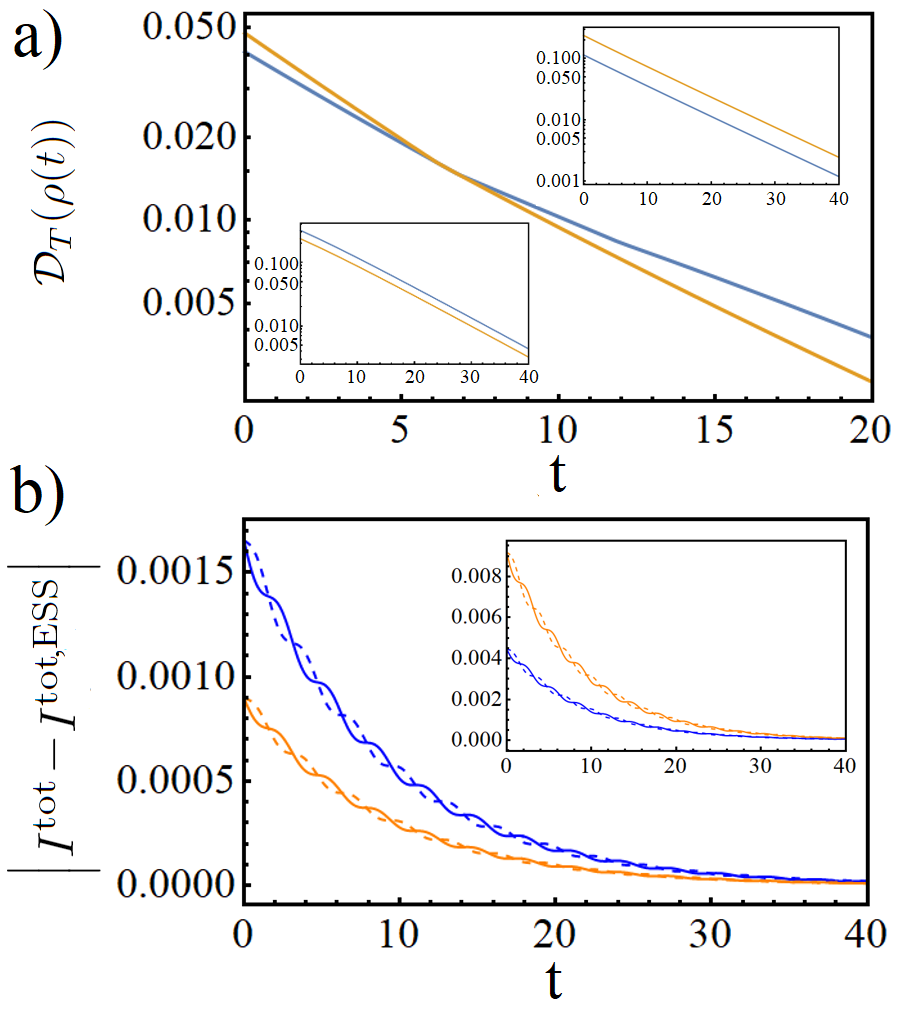}
\end{center}
\caption{Examples for QMEs for the I2KM with parameters as specified in Fig.~\ref{fig3}. 
Yellow [blue] curves refer to the ``far'' [``close''] initial state $\rho_f(0)$ [$\rho_{c}(0)$] as defined by ${\cal D}_E$ in Eq.~\eqref{Eucldist}.
\emph{Panel (a):} ${\cal D}_T$ (in log scale) vs $t$.
Main panel: type-II QME, where $(T_{L,c}^{-1}, \mu_{L,c})=(0.7,2.7)$ and $(T_{L,f}^{-1}, \mu_{L,f})=(0.7,3)$.
Lower-left inset: type-I QME, with $(T_{L,c}^{-1},\mu_{L,c})=(1.3,1.5)$ and $(T_{L,f}^{-1},\mu_{L,f})=(0.7,1.5)$.
Upper-right inset: no QME, where $(T_{L,c}^{-1},\mu_{L,c})=(1.7,2.7)$ and $(T_{L,f}^{-1},\mu_{L,f})=(1.7,3)$.
\emph{Panel (b):} $\left|I_\lambda^{\rm tot}(t)-I^{\rm tot, ESS}\right|$ vs $t$, 
see Eq.~\eqref{current}.  Solid (dashed) curves correspond to currents from lead 
$\lambda=L$ ($\lambda=R$).
Main panel: type-I QME, with parameters as 
in the lower-left inset of panel (a).  Inset: no QME, cf.~the upper-right inset of panel (a).
}
    \label{fig4}
\end{figure}

To decide which type of QME occurs, one needs to analyze the time dependence of the trace distance ${\cal D}_T$ in Eq.~\eqref{tracedist}.
As shown in Fig.~\ref{fig4}(a), depending on the parameter regime, one can either realize a type-I QME or the more elusive type-II QME.  
The latter is characterized by an intersection point at a finite time $t^*$.
We further discuss this point in the SM \cite{SM}, where we compare the isolines of the trace distance \eqref{tracedist} at $t=0$ to those of the Euclidean distance \eqref{Eucldist}.
However, in order to obtain curves such as those in Fig.~\ref{fig4}(a) in experiments, one has to perform quantum state tomography. A much less costly way to identify the QME is to measure the time-dependent currents \eqref{current}, see Fig.~\ref{fig4}(b).
Alternatively, one can also the quantities in Eq.~\eqref{current2} \cite{SM}. However, a measurement of the current \eqref{current} or of the observables in  
Eq.~\eqref{current2} is \emph{not} sufficient to differentiate between 
type-I and  type-II QMEs. The latter distinction requires state tomography.

\emph{Conclusions.---}We have formulated a general and widely applicable protocol which allows one to unambiguously identify and classify the QME in open nonequilibrium quantum systems.  Our approach shows that quantum correlations are of fundamental importance for the predicted phenomena. As practical example, we have applied our ideas to a minimal interacting two-site Kitaev model. Our theory can be readily generalized to more complicated open topological systems  \cite{Nava2023,Cinnirella_2024,Nava_2023,Nava_2024}. While the distinction of type-I and type-II QMEs in general requires quantum state tomography, one can infer the existence of a QME already by monitoring the time dependence of simpler observables such as the electrical current.  

\begin{acknowledgments} 
We thank M. Fabrizio and D. Giuliano for discussions. We
 acknowledge funding by the Deutsche Forschungsgemeinschaft (DFG, German Research Foundation) under Projektnummer 277101999 - TRR 183 (project B02), under Project No.~EG 96/13-1, and under Germany's Excellence Strategy - Cluster of Excellence Matter and Light for Quantum Computing (ML4Q) EXC 2004/1 - 390534769.
\end{acknowledgments}

{\color{black}
\section{Supplemental Material}
\label{SM}

We here provide further details on our work.  In Sec.~\ref{sec1}, we comment on
previous theories for the quantum Mpemba effect (QME) in open nonequilibrium systems which suggest to employ finite-time crossing points
of certain observables for identifying the QME. We also compare their predictions to those made by our protocol. 
In Sec.~\ref{sec2}, we provide further details on  experimentally accessible quantities 
for the interacting two-site Kitaev model (I2KM) discussed in the main text. In Sec.~\ref{sec3}, we comment
on the optimal working conditions and on the role of the geometric angle $\theta$ of the I2KM. 
In Sec.~\ref{sec4}, we discuss the sweet parameter spot of the I2KM, where one has fine-tuned Majorana bound states. Finally, in Sec.~\ref{sec5}, we provide a detailed discussion of distance measures. 
For brevity, Eq.~(X) in the main text is referred to as Eq.~(MX) below.

\section{I. Crossing points}\label{sec1}

In recent theoretical work \cite{Chatterjee2023,Chatterjee2023_2,Wang2024}, the QME has been linked to finite-time crossings of temporal trajectories of 
certain observables.  Such finite-time crossings are searched for in various physical quantities, using two system copies with different initial (hot vs cold) parameters but the same after-quench (N)ESS parameters.
Typical observables include the diagonal elements  of the system density matrix (``populations''), an effective temperature, the von Neumann
entropy, and the average system energy. However, all those quantities are neither bounded from above or below, and they are generally not monotonically decreasing (or increasing) functions of time. For this reason, they can cause  false detection of the QME or overlook the QME when it is
actually present.  We note that related arguments have already been given in 
Ref.~\cite{Lu2017} for the classical (thermal) equilibrium ME. 
In the following, we argue that the existence of one or multiple crossing points is neither a necessary nor a sufficient condition for the onset of the QME.

In Ref.~\cite{Chatterjee2023}, a spinful interacting quantum dot (ISD) coupled to two leads has been considered. The total Hamiltonian,
$H=H_{\rm{ISD}}+H_{l}+H_{c}$, contains the ISD Hamiltonian,
\begin{equation} \label{one-site}
H_{\rm{ISD}}= \sum_\sigma \epsilon\,n_\sigma + U n_\uparrow n_\downarrow,
\end{equation}
where $c_\sigma$ and $n_\sigma=c_\sigma^{\dagger} c_\sigma^{}$ are electron annihilation and occupation operators for spin $\sigma\in\{\uparrow,\downarrow\}$,
respectively. Here $\epsilon$ is a single-particle energy and $U$ the Coulomb charging energy. 
Below, we use the many-body basis 
\begin{equation}\label{many-body-one-site}
    \{ \ket{0}, \, \ket{\uparrow}=c_\uparrow^\dagger|0\rangle,\, \ket{\downarrow}=c_\downarrow^\dagger|0\rangle,\, \ket{d}= c_\uparrow^{\dagger}c_\downarrow^{\dagger}\ket{0} \},
\end{equation}
which diagonalizes the Hamiltonian \eqref{one-site}.  
The Hamiltonian $H_{l}$ describes the left and right ($\lambda=L,R$) leads as spin-degenerate ideal Fermi gases, similar to the main text. 
The tunneling Hamiltonian $H_{c}$ connecting the leads and the ISD is given by
\begin{equation} \label{one-site-bath}
H_{c}=\sum_{\lambda,k,\sigma} t_{\lambda} c_{\lambda,k,\sigma}^{\dagger} c_{\sigma}^{} +\textrm{h.c.},
\end{equation}
where $c_{\lambda,k,\sigma}$ is the electron annihilation operator for lead $\lambda$, momentum $k$, and spin $\sigma$. Following Ref.~\cite{Chatterjee2023}, the tunneling amplitude $t_{\lambda}$ between lead $\lambda$ and the ISD is assumed spin- and energy-independent. 

In the wide-band approximation, and using the Born-Markov and rotating wave approximations, one can integrate out the leads and derive a Lindblad 
master equation for the density matrix $\rho(t)$ of the ISD model.  We express $\rho$ as $4\times 4$ matrix in the basis \eqref{many-body-one-site}.   
In contrast to the I2KM, the dissipative jump operators mix \emph{all} system states for the ISD model.
As a consequence, the dynamical equations for the diagonal elements of the density matrix (i.e., the populations),
\begin{equation}\label{popul}
{\bf P}(t)  \equiv \left(\begin{array}{c}\rho_{0,0}(t) \\ \rho_{\uparrow,\uparrow}(t)\\ \rho_{\downarrow,\downarrow}(t)\\ \rho_{d,d}(t)\end{array}\right),
\end{equation}
decouple from the off-diagonal entries of $\rho(t)$. (For simplicity, we assume that one starts initially with a diagonal state $\rho(0)$.)
One then arrives at a \emph{Pauli master equation} for the populations \cite{breuer2007theory}, 
\begin{equation}\label{pauli}
\partial_t {\bf P}(t) = {\bf M} \, {\bf P}(t),
\end{equation}
where ${\bf M}$ is given by the $4 \times 4$ matrix
\begin{equation}
\label{pauliM} 
\left(\begin{array}{cccc}
-\sum_\sigma\Gamma_{\sigma,0} & \Gamma_{0,\uparrow} & \Gamma_{0,\downarrow} & 0\\
\Gamma_{\uparrow,0} & -\Gamma_{d,\uparrow}-\Gamma_{0,\uparrow} & 0 & \Gamma_{\uparrow,d}\\
\Gamma_{\downarrow,0} & 0 & -\Gamma_{d,\downarrow}-\Gamma_{0,\downarrow} & \Gamma_{\downarrow,d}\\
0 & \Gamma_{d,\uparrow} & \Gamma_{d,\downarrow} & -\sum_\sigma\Gamma_{\sigma,d}
\end{array}\right).
\end{equation}
We use the rates 
\begin{eqnarray}\label{pauliRates}
    \Gamma_{\sigma,0}&=&\gamma \sum_{\lambda=L,R} f_\lambda(\epsilon)= 2\gamma- \Gamma_{0,\sigma}, \nonumber \\
    \Gamma_{d,\sigma}&=&\gamma \sum_{\lambda}  f_\lambda(\epsilon+U) = 2\gamma- \Gamma_{\sigma,d},
\end{eqnarray}
where $f_\lambda(\epsilon)$ is the Fermi function and $\gamma=2\pi \nu_\lambda(\epsilon) |t_\lambda|^2$
as in the main text.  The microscopic transition rate $\gamma$ is assumed to be independent of energy, spin, and the lead index \cite{Chatterjee2023}.

\begin{figure}
\begin{center}
    \includegraphics[width=0.5\textwidth]{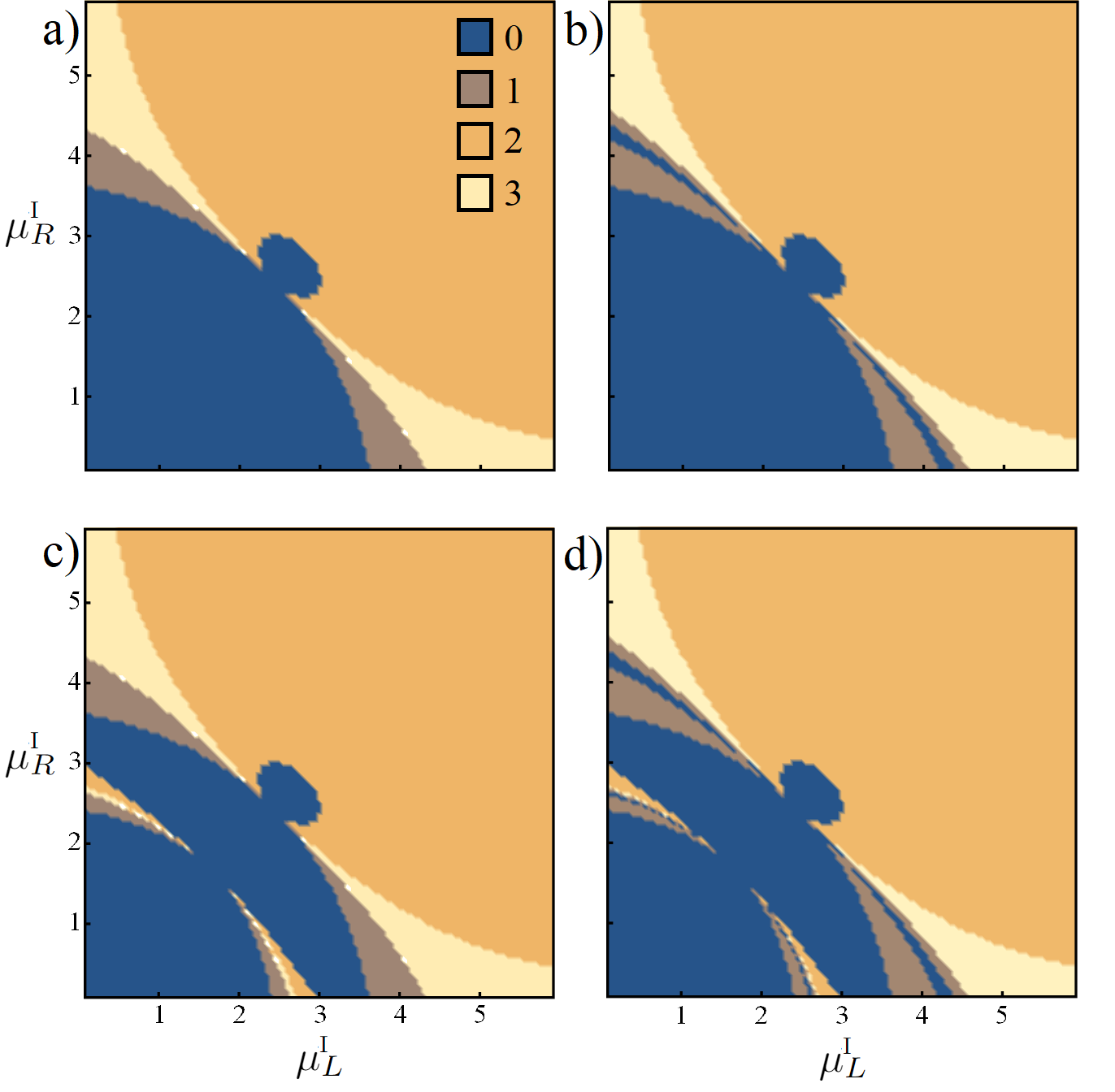}
    \caption{Phase diagram for the ISD model in the $\mu_R^I$-$\mu_L^I$ plane for otherwise fixed parameters, cf.~Ref.~\cite{Chatterjee2023}. 
    Two copies of the system are initially (at time $t=0$) prepared in the equilibrium density matrix corresponding to $(\mu_L^{I}, \mu_R^{I}$) with $\beta_L^{I}=\beta_R^{I}=1.15$, and $\mu_L^{II}=\mu_R^{II}=2.43$ with $\beta_L^{II}=\beta_R^{II}=1.15$, respectively. After the quench, the left and right leads have chemical potentials $\mu_L^{\rm ESS}=\mu_R^{\rm ESS}=2$ and inverse temperatures $\beta_L^{\rm ESS}=\beta_R^{\rm ESS}=1$.  The other parameters are set to $U=1.25$, $\epsilon=2$, and $\gamma=1$.
    {\bf (a):} Number of populations $P_{\eta}$, see Eq.~\eqref{popul}, exhibiting a crossing point for a finite time $t^*$ as a function of $\mu_L^{I}$ and $\mu_R^{I}$. {\bf (b):}  Same as panel (a) but applying a cut-off scheme to account for a finite measurement resolution.  Here, crossing points are discarded if
    $\left|P_\eta^{I/II}(t^*)-P_\eta^{\rm ESS}\right| <10^{-5}$. 
    {\bf (c):} Crossing points are now instead searched for by analyzing $\left|P_\eta(t)-P_\eta^{\rm ESS}\right|$, see Ref.~\cite{Chatterjee2023}.
    {\bf (d):} Same as in panel (c) but with the analogous cut-off procedure as applied in panel (b).}
    \label{figs1}
\end{center} 
\end{figure}

In Ref.~\cite{Chatterjee2023}, the following protocol to search for the QME has been introduced. Two copies of the system (labeled by $I$ and $II$, respectively) 
are prepared in the equilibrium configuration ${\bf P}^{I(II)}(0)$, defined by the chemical potentials and inverse temperatures of both leads, 
$\left\{\mu_L^{I(II)},\mu_R^{I(II)},\beta_L^{I(II)},\beta_R^{I(II)}\right\}$. 
(We emphasize that the indices $I/II$ have nothing to do with the type-I or type-II QME introduced in the main text. Instead, $I$ and $II$ here refer to different initial configurations.)
At time $t=0$, the lead parameters are quenched to their final ESS values 
$\left\{\mu_L^{\rm ESS},\mu_R^{\rm ESS},\beta_L^{\rm ESS},\beta_R^{\rm ESS}\right\}$, and the subsequent time evolution of ${\bf P}^{I(II)}(t)$ under the Pauli equation \eqref{pauli} is monitored.
The (first) crossing point $t^*$ is defined by the condition $P_\eta^I(t^*)= P_\eta^{II}(t^*)$ for some component of the population vector \eqref{popul}. In Fig.~\ref{figs1}(a), cf.~Fig.~1 in Ref.~\cite{Chatterjee2023}, 
we show the number of population components that exhibit a crossing point as a function of $\mu_L^I$ and $\mu_R^I$ for an otherwise fixed parameter set.  
One finds different phases characterized by a different number of population components having a crossing point. The authors of Ref.~\cite{Chatterjee2023} claim that these crossing points correspond to the onset of the QME for the corresponding observable.
However, in general, these density matrix elements are neither monotonically decreasing nor increasing functions of time. Indeed, in general, they may approach their final (N)ESS value from above, from below, or in an oscillatory manner. The oscillatory behavior is excluded for the Pauli equation since the Hamiltonian does not affect Eq.~\eqref{pauli}. For the ISD model, the time evolution is thus purely exponential. However, oscillations can occur in other setups like the I2KM, where jump operators couple states that are not eigenstates of the Hamiltonian.

\begin{figure}
\begin{center}
    \includegraphics[width=0.33\textwidth]{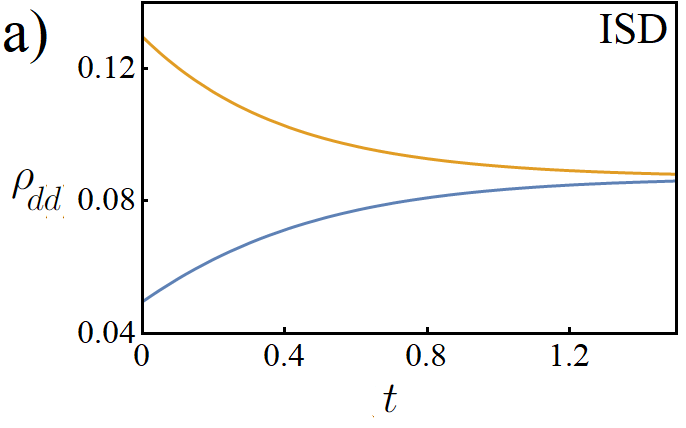}
    \includegraphics[width=0.33\textwidth]{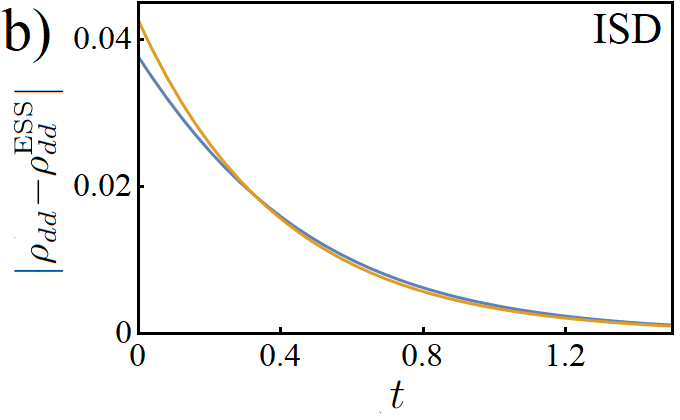}
    \includegraphics[width=0.33\textwidth]{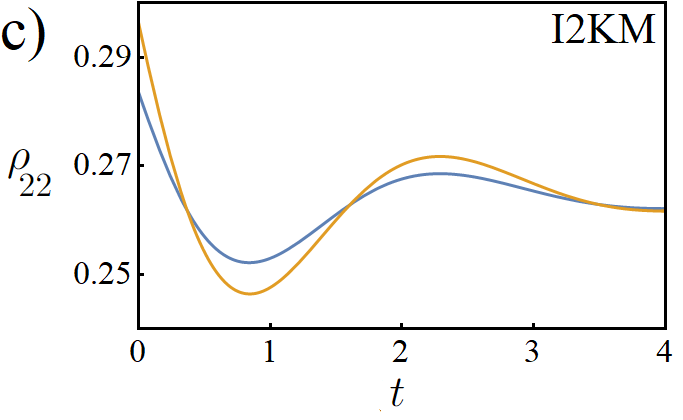}
    \caption{On crossing points in the time domain. {\bf (a):} Time evolution of $\rho_{d,d}^I(t)$ (blue) 
    and $\rho_{d,d}^{II}(t)$ (yellow curve) for the ISD model, using the parameters in Fig.~\ref{figs1} 
    with $\mu_L^I=2.5$ and $\mu_R^I=0.1$.  {\bf (b):} Time evolution of $\left|\rho_{d,d}^I(t)-\rho_{d,d}^{\rm ESS}\right|$ (blue) and 
    $\left|\rho_{d,d}^{II}(t)-\rho_{d,d}^{\rm ESS}\right|$ (yellow) for the ISD model, using again the parameters in panel (a).  
    {\bf (c):} Time evolution of $\rho_{2,2}^I(t)$ (blue) and $\rho_{2,2}^{II}(t)$ (yellow) for the I2KM 
    with $t_h=1$, $\epsilon_1=\epsilon_2=2$, $\Delta=0.1$, $U=0.25$, $\Gamma=0.1$, and $\theta=0$. The system is driven towards a final ESS with $\mu_{L}^{\rm ESS}=\mu_{R}^{\rm ESS}=2.4$ and $\beta_{L}^{\rm ESS}=\beta_{R}^{\rm ESS}=2.15$, starting from two different pre-quench configurations with the same parameters except for $\mu_{L}^I=\mu_L^{II}=1.5$, $\beta_{L}^I=1.3$,
    and $\beta_{L}^{II}=0.7$.}
    \label{figs2}
\end{center} 
\end{figure}

We observe from Fig.~\ref{figs1} that the occurrence or absence of the QME depends on the monitoring function that one employs. In particular, using either the populations, see panel (a), or the absolute values of the population deviations from their final values, 
see panel (c), gives different phase diagrams.  Moreover, since here we do not have a phase transition like in the original work on the classical ME \cite{Mpemba_1969}, the system reaches the final (N)ESS only after infinitely long time through an exponential decay \cite{Lu2017}. In practice, the critical time $t^*$ at which the crossing takes place can then become very large. At the same time, 
the distance of the monitored observable from its (N)ESS value then becomes extremely small, beyond numerical or experimental precision. For this reason, it is important to compare panels (a,c) with a situation where one imposes a resolution limit 
on the distance measurement.  Once this limit has been reached, the search for a crossing point is terminated.
In Fig.~\ref{figs1}(b,d), we use such a cut-off scheme in order to implement this consideration. 
Evidently, the resulting phase diagrams in panels (b,d) are substantially different from the corresponding ones 
in panels (a,c), which were obtained by assuming perfect resolution capabilities.

Next, in Fig.~\ref{figs2}(a), we show the time evolution of $\rho_{d,d}^I(t)$ and $\rho_{d,d}^{II}(t)$ for the parameters in Fig.~\ref{figs1} with $\mu_L^I=2.5$ and $\mu_R^I=0.1$. 
No crossing point is detected at any finite time but, clearly, the two density matrix elements approach the equilibrium value from above (initial condition $II$) and below (initial condition $I$). However, by comparing the time dependence of $\left|\rho_{d,d}^I(t)-\rho_{d,d}^{\rm ESS}\right|$ and $\left|\rho_{d,d}^{II}(t)-\rho_{d,d}^{\rm ESS}\right|$, see Fig.~\ref{figs2}(b), we find
that a critical time $t^*$ exists after which $\rho_{d,d}^{II}(t>t^*)$ is nearer to the final ESS value than $\rho_{d,d}^I(t>t^*)$. Together with Fig.~\ref{figs1} and the corresponding discussion, those observations 
show that the existence of a crossing point is in general \emph{not} a necessary condition for the QME to take place.

Let us next show an example where the existence of crossing points does not give a \emph{sufficient} condition for the QME. 
In the presence of both Lindblad and Liouvillian time evolution, as is the case for the I2KM, 
an oscillatory behavior is observed in many quantities. 
For such models, many crossing points can emerge, while no true QME is detected from
the envelope of the oscillatory functions (for these parameters). In fact, 
in Fig.~\ref{figs2}(c), for the I2KM case, we show the time evolution of the population of the  state $\ket{2}$,
see Eqs.~(M4) and (M5), after two different initial conditions.  Note that this state is not an eigenvector 
of the Hamiltonian. Evidently, the oscillatory behavior 
leads to finite-time crossings which artificially suggest a QME while the envelopes show that 
one of the curves is actually always closer to the final (N)ESS value. 

To summarize, finite-time crossings of density matrix elements \cite{Chatterjee2023,Chatterjee2023_2,Wang2024}
do not provide a reliable guide to the identification of the QME.
Crossing points can emerge due to oscillatory behavior, or they can simply be absent because the density matrix elements, for different initial conditions, approach the (N)ESS from above and below, respectively.
It is worth mentioning that the presence of a crossing point in a generic observable could be purely accidental, e.g., due to natural constraints that the system must satisfy.  For populations, for example,
we have $\sum_i \rho_{i,i}(t) =1$ due to state normalization. 
As acknowledged in Ref.~\cite{Wang2024}, quantum correlations are of fundamental importance for the QME. However, density matrix elements do not provide a suitable observable in this context.

\begin{figure}
\begin{center}
    \includegraphics[width=0.35\textwidth]{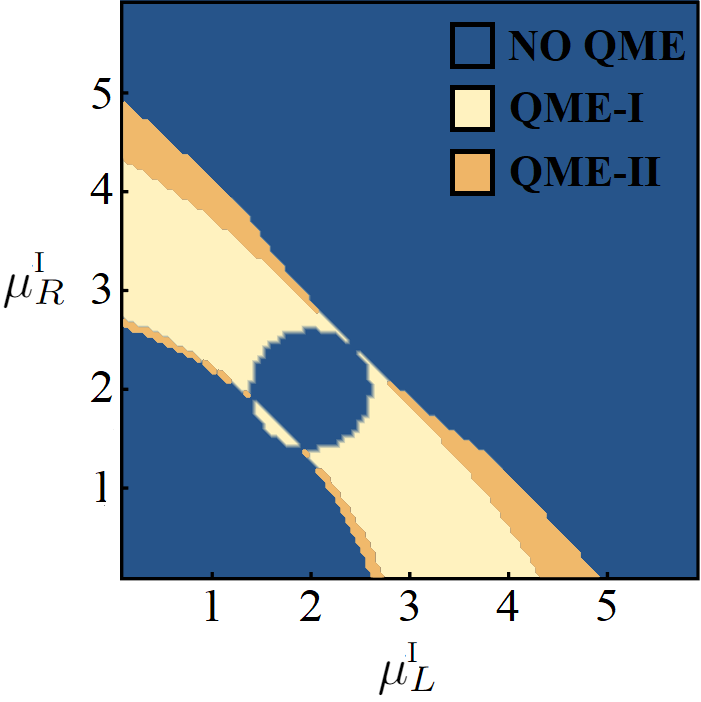}
    \caption{Phase diagram for the ISD model in the $\mu_R^I$-$\mu_L^I$ plane, obtained by using our trace distance protocol for the same parameters as in Fig.~\ref{figs1}(c).  
    Type-I and type-II QMEs (here dubbed QME-I and QME-II) have been identified as explained in the main text.}
    \label{figs3}
\end{center} 
\end{figure}

We next compare the phase diagram obtained from the crossing point procedure, see in particular Fig.~\ref{figs1}(c),
with the corresponding phase diagram  for the ISD model computed from the trace distance protocol proposed in the main text.   
In Fig.~\ref{figs3}, we show our results, which were obtained for the same parameters as in Fig.~\ref{figs1}(c). 
It is rather obvious that the corresponding phase diagrams in Fig.~\ref{figs1}(c) and Fig.~\ref{figs3} are very different. 
We conclude that,  although it is legitimate to compare the time evolution of a generic system observable for different initial conditions, the existence of crossing points in the time evolution of density matrix elements
is not one-by-one related to a QME.  In our protocol, see Fig.~\ref{figs3} for the ISD model, the QME is instead detected through properly defined distance-from-equilibrium measures. This procedure is also able to distinguish type-I and type-II QMEs.

\section{II. Experimental quantities}\label{sec2}  

In the main text, we introduced the trace distance ${\cal D}_T$ in Eq.~(M2) as a proper distance measure  
for quantum states $\rho(t)$ from the final (N)ESS. The trace distance allows one to unambiguously detect the existence of the QME and its type.
Moreover, it does not suffer from the limitations imposed by 
many other observables, as discussed in Sec.~\ref{sec1}. However, the trace distance is not an easily measurable quantity as it requires quantum state tomography. While this task is in principle possible, it is a challenge in practice.   For this reason, we here compare its behavior with other quantities that can be experimentally measured with less effort. We focus on the I2KM introduced in the main text.

\begin{figure}
\begin{center}
    \includegraphics[width=0.45\textwidth]{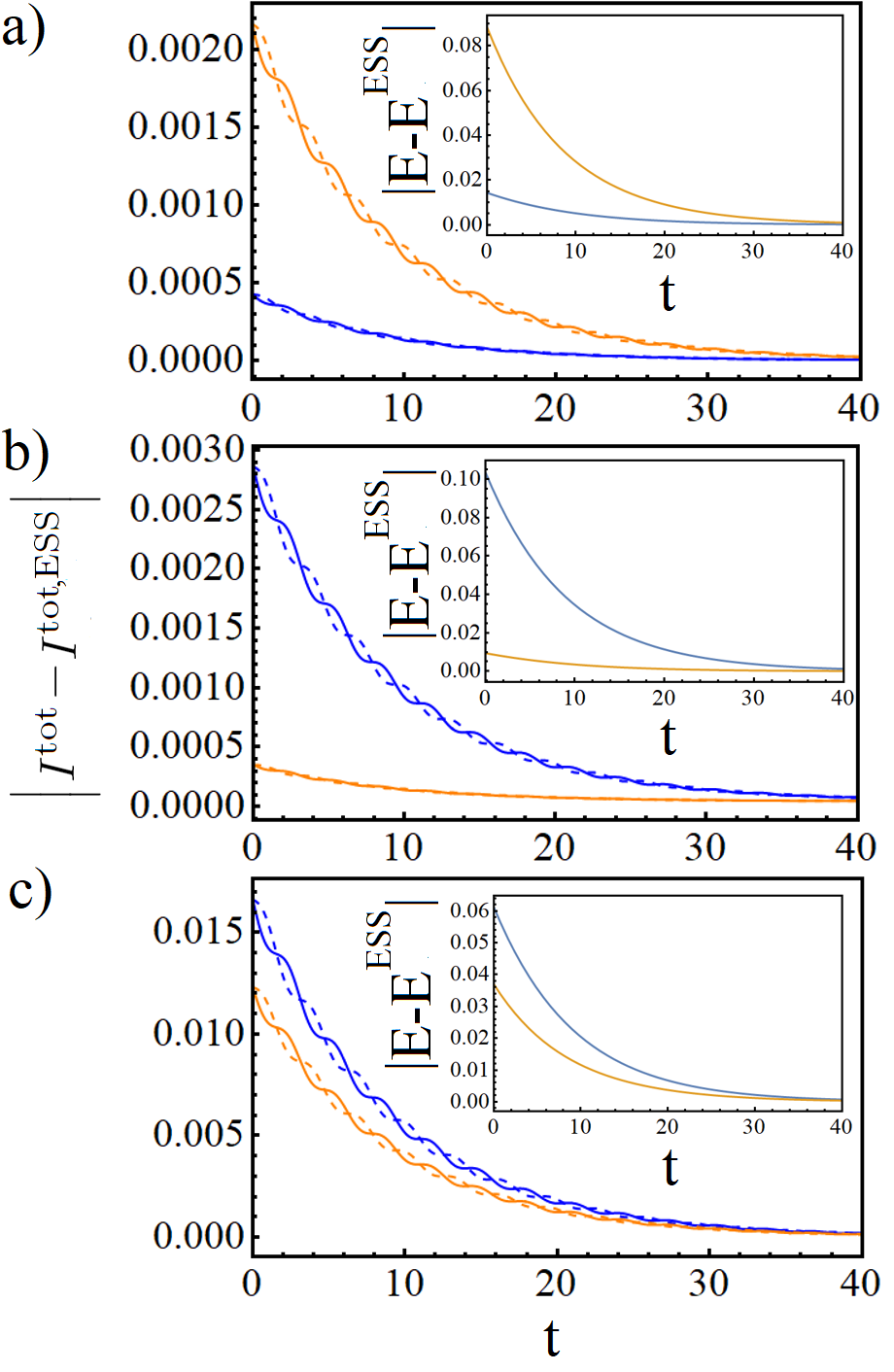}
    \caption{Currents ($I_\lambda^{\rm tot}$) and energy ($E$) in the time domain for the I2KM, where the indices $I$ vs $II$ refer to ``close'' vs ``far'' initial conditions. Main panels: Time evolution of current differences $\left|I_{\lambda}^{{\rm tot},I}(t)-I^{\rm tot, ESS}\right|$ (blue) and 
    $\left|I_{\lambda}^{{\rm tot},II}(t)-I^{{\rm tot, ESS}}\right|$ (yellow), for the current flowing from the left ($\lambda=L$, solid) and right ($\lambda=R$, dashed) lead to the I2KM, respectively. We set $(T_{L,I}^{-1}, \mu_{L,I})=(0.7,2.7)$ and $(T_{L,II}^{-1}, \mu_{L,II})=(0.7,3)$. Others parameters are as in Fig.~3 of the main text, 
    except for $\epsilon_1=\epsilon_2=\epsilon$ as specified below.
    {\bf (a):} Case without QME, where $\epsilon=2.08$.
    {\bf (b):} Type-I QME ($\epsilon=1.92$). {\bf (c):}  Type-II QME ($\epsilon=2$).  Insets: Time evolution of energy differences $\left|E^I(t)-E^{\rm ESS}\right|$ (blue) and $\left|E^{II}(t)-E^{\rm ESS}\right|$ (yellow) to the final state for the two initial states, respectively.}
    \label{figs4}
\end{center} 
\end{figure}

In Fig.~\ref{figs4}, we show the time evolution of the total current $I_\lambda^{\rm tot}(t)$ exchanged between lead $\lambda\in\{L,R\}$ and the I2KM, see Eq.~(M10), and the time evolution of the energy $E(t)$ of the I2KM, see Eq.~(M11).  In the main panels, we plot the absolute value of the difference of the currents with respect to their final ESS values 
(which are independent of $\lambda$).
When no QME is revealed by the trace distance protocol, the current of the ``close'' system (initial condition $I$)  
remains nearer to the ESS value for all times $t$ than the current of the ``far'' one (initial condition $II$), see the main panel in 
Fig.~\ref{figs4}(a).  
On the other hand, when a type-I or a type-II QME is revealed by the trace distance protocol, cf.~Fig.~4(a) in the main text, the distance between the time-dependent current and the ESS value of the ``far'' system remains always below the one for 
the ``close'' system, see the main panels in Fig.~\ref{figs4}(b,c), respectively. 
However, no crossing point is detected in Fig.~\ref{figs4}(c), and therefore the more elusive type-II QME  cannot 
be distinguished from the type-I QME by simply analyzing the current. 
We conclude that quantum state tomography seems necessary to differentiate between type-I and type-II QMEs.

Very similar conclusions can also be drawn by analyzing the time-dependent energy of the I2KM system.  For the three relevant cases (no QME, type-I QME, type-II QME), the corresponding curves for $\left| E(t)-E^{\rm ESS}\right|$ are shown in the respective insets of Fig.~\ref{figs4}.  Again, it is not possible to distinguish type-I and type-II QMEs based on this observable.  However, one can distinguish the presence or absence of a QME in an unambiguous manner by measuring this observable.

In fact, the existence of a QME --- leaving aside the classification into type-I or type-II cases --- 
can already be established by measuring the currents $I_\lambda^{\rm tot}(t)$ or the energy $E(t)$ (relative to their final ESS values)
at a time $t=0^+$ shortly after the parameter quench. The presence of a QME is characterized by the condition 
\begin{equation}
    \left|I_{\lambda}^{{\rm tot},I}(0^+)-I_{}^{\rm tot, ESS}\right|> \left|I_\lambda^{{\rm tot},II}(0^+)-I_{}^{\rm tot, ESS}\right|,
\end{equation}
while otherwise no QME occurs. A similar condition applies to the total energy of the I2KM. 
The current between the two dots, $I_{1,2}$ in Eq.~(M11), also allows one to draw the same conclusions in principle. 
However, $I_{1,2}$ is a strongly oscillatory function of time, since it also depends on the off-diagonal elements of $\rho(t)$.  Nonetheless, its envelope, under initial condition $I$, always stays further away from (closer to) the final ESS value 
than for initial condition $II$ if the QME is present (absent). On the contrary, finite-time crossings of density-matrix based quantities like the von Neumann entropy or populations may result in misleading predictions, see Sec.~\ref{sec1}.  
Such pitfalls are avoided by following our protocol.

\begin{figure}
\begin{center}
    \includegraphics[width=0.45\textwidth]{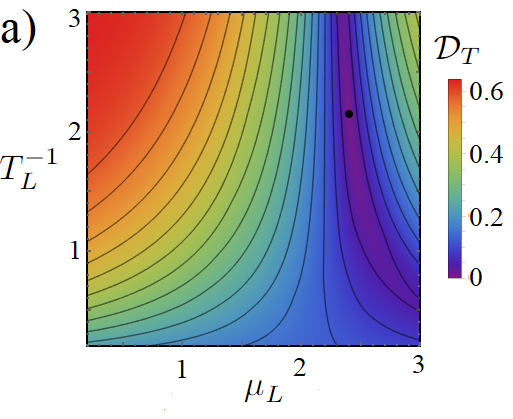}
    \includegraphics[width=0.45\textwidth]{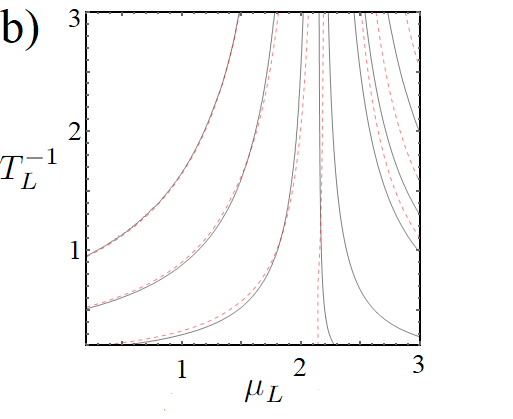}
    \caption{{\bf (a):} Color-scale plot of the I2KM trace distance $\mathcal{D}_T (\rho (0))$ with respect to the final ESS (indicated by a black circle) in the $T_L^{-1}$-$\mu_L$ plane for different initial NESS configurations $(T^{-1}_L,\mu_L)$, where $(T^{-1}_R,\mu_R)$ is identical to the final ESS parameters, $T^{-1}_{\lambda,{\rm ESS}}=2.15$ and $\mu_{\lambda,{\rm ESS}}=2.4$ for $\lambda=L,R$. Other parameters are $t_h=1$, $\epsilon_1=\epsilon_2=2$, $\Delta=0.1$, $U=0.25$,  $\Gamma=0.1$, and $\theta=0$. Thin black curves indicate $\mathcal{D}_T$-isolines. {\bf (b):} Comparison between $\tau$-isolines [black solid lines, cf.~Fig.~3 of the main text] and $\mathcal{D}_T$-isolines [red dashed lines, cf.~panel (a)]. A deviation between 
    these two isoline sets  indicates a type-II QME.}
    \label{figs5}
\end{center} 
\end{figure}

If quantum state tomography is available, one can directly implement the trace distance protocol as discussed in the main text. In Fig.~\ref{figs5}, we show that a comparison of the trace distance ${\cal D}_T$ between the close and far initial states and the final ESS with the respective values of the Euclidean distance ${\cal D}_E$ can directly reveal a type-II QME. Figure~3 in the main text shows how the absence or presence of the QME can be detected from the $\tau$-isolines in the bath parameter space. 
However, this information is not sufficient to discriminate between type-I and type-II QME. For that purpose, one needs to 
check the condition $\mathcal{D}_T (\rho_f (0)) \gtrless \mathcal{D}_T (\rho_c (0))$ for the trace distance, see Fig.~\ref{figs5}(a). Choosing, e.g., the ``close'' initial condition, 
we analyze the corresponding $\tau$-isoline and the $\mathcal{D}_T$-isoline crossing at that point. If there is a mismatch between the two isolines, see Fig.~\ref{figs5}(b), a small region for the onset of the QME-II exists. In this region, the relaxation time starting with $\rho_c(0)$ is longer than if one starts from $\rho_f(0)$, but the trace distance for $\rho_c(0)$ is still smaller than for $\rho_f(0)$.

\section{III. Optimal working point}\label{sec3}

In this section, we point out an interesting behavior that, albeit not directly connected with the QME, could play an important role in systems with complex geometries and in experiments. 
As observed in previous works \cite{Benenti2009,Nava2021}, when an interacting one-dimensional electronic chain is connected to two reservoirs, an optimal working point emerges through a change in the monotonicity of the NESS current as a function of the coupling between the chain and the reservoirs. 
The optimal working point is a consequence of the presence of two time scales in the dissipative quantum dynamics.
First, there is an intrinsic time scale induced by the Hamiltonian. Second, there is a dissipative time scale set by the coupling strength $\Gamma$ to the baths. 

For the I2KM, this effect appears through the geometric angle $\theta$ in Eq.~(M9). For $\theta=0$, the system is connected in series with the two leads and the current $I_{1,2}$ between both dots is maximal. Indeed, an electron incoming from the left lead is forced to jump from dot $1$ to dot $2$ in order to reach the right lead. On the contrary, for $\theta=\pi/2$, the system is connected in parallel with the leads, and an electron can move from left to right by just populating one of the two dots. 
In this case, the classical vs quantum competition is reduced, with a qualitatively different dependence on $\Gamma$. 

\begin{figure}
\begin{center}
    \includegraphics[width=0.45\textwidth]{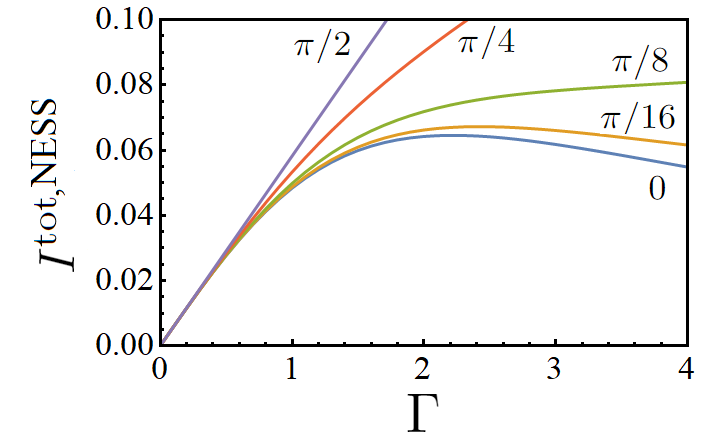}
    \caption{The NESS current $I^{\rm tot, NESS}$ (in units of $et_h/\hbar$) of the I2KM vs system-lead hybridization strength $\Gamma$, see Eq.~(M10), for $t_h=1$, $\epsilon_1=\epsilon_2=2$, $\Delta=0.1$, and $U=0.25$. 
     For arbitrary initial states, the system is driven towards a final NESS with $\mu_{L}^{\rm NESS}=2.4$, $\mu_{R}^{\rm NESS}=2.2$, $\beta_{L}^{\rm NESS}=2.15$, and $\beta_{R}^{\rm NESS}=1.15$. We study different geometric angles $\theta$ as specified near each curve. }
    \label{figs6}
\end{center} 
\end{figure}

In Fig.~\ref{figs6}, the NESS current $I^{\rm tot, NESS}$, which neither depends on the chosen initial state nor on the lead index $\lambda$, is shown as a function of $\Gamma$ for different  $\theta$.  We observe that the largest current is obtained for $\theta=\pi/2$, while for $\theta=0$, the current has a maximum at a finite hybridization $\Gamma$, with a linear increase of the current for small $\Gamma$.  An optimal working point thus emerges for $\theta=0$, or more generally for small values of $\theta\alt \pi/16$.
In more complex geometries, such effects should be taken into consideration as they could affect not only the NESS current \cite{Benenti2009,Nava2021} but also the relaxation time towards the final NESS configuration \cite{Artiaco2023}.

\section{IV. Sweet spot of I2KM}\label{sec4}

At the sweet parameter spot of the I2KM, $\Delta=t_h=1$ and $U=\epsilon_1=\epsilon_2=0$, two fine-tuned ``poor man's" Majorana bound states are localized on the first and second dot, respectively, without mutual overlap.  This effectively cuts the I2KM into two halves, resulting in a vanishing current $I_{1,2}=0$. For $\theta=0$, 
the two leads are therefore dynamically decoupled. Here each dot relaxes to its own equilibrium state, which depends on the temperature and the chemical potential of the lead to which it is physically coupled. 
On the other hand, for $\theta=\pi/2$, a current can flow between both leads (but not between the dots), and the system effectively reduces to two independent dots coupled to both leads.  

\begin{figure}
\begin{center}
    \includegraphics[width=0.45\textwidth]{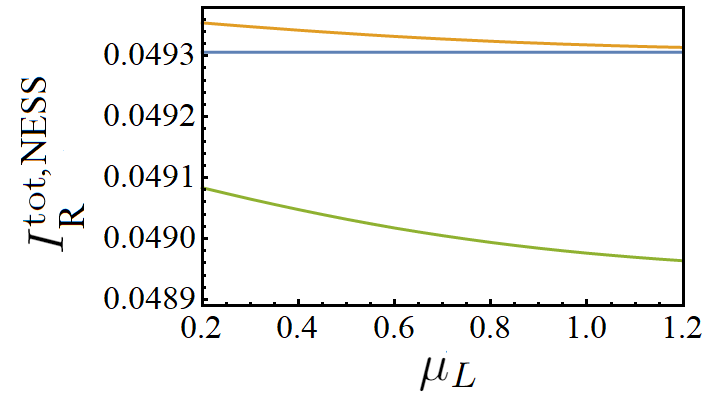}
    \caption{Sweet spot of the I2KM and effects of fine-tuned Majorana bound states. The NESS current $I_R^{\rm tot, NESS}$ between the I2KM and the right lead (in units of $et_h/\hbar$) is shown as function of 
 $\mu_L=\mu_L^{\rm NESS}$ for $U=\epsilon_1=\epsilon_2=0,\, t_h=1, \, \Gamma=0.1, \, \mu_{R}^{\rm NESS}=2.4,\, \beta_{L}^{\rm NESS}=2.15$, and $\beta_{R}^{\rm NESS}=2.15$. The blue curve is for $\Delta=1$ and $\theta=0$, the yellow curve for $\Delta=1$ and $\theta=\pi/2$, and the green curve for $\Delta=1/2$ (away from the sweet spot) and $\theta=0$. 
 }  
    \label{figs7}
\end{center} 
\end{figure}

In Fig.~\ref{figs7}, we consider the sweet spot regime and its vicinity. We show the NESS current $I_R^{\rm tot, NESS}$, see Eq.~(M10), between the I2KM and the right lead
as a function of $\mu_L=\mu_L^{\rm NESS}$ for fixed $\mu_R^{\rm NESS}$.  For $\theta=0$, the current between the right dot and the right lead is not affected by the left dot, and thus stays constant for all $\mu_L$. 
At the same time, the non-zero NESS current is a signature for the poor man's Majorana bound states which convert the incoming dissipative current into a supercurrent by means of Andreev reflection processes. 
For $\theta=\pi/2$, $I_R^{\rm tot, NESS}$ instead depends on $\mu_L$  due to a now finite transmission probability and the corresponding crossed Andreev reflection amplitude. 
Detuning the Hamiltonian parameters from the sweet spot condition, e.g., by setting $\Delta=1/2$ (see the green curve in Fig.~\ref{figs7}),
will destroy the poor man's Majorana states. In such a case, the system is not anymore cut in two halves, and $I_R^{\rm tot, NESS}$ depends on $\mu_L$ even for $\theta=0$. 
At the same time, the current is reduced (compared to $\Delta=1$) due to the smaller Andreev reflection probability previously mediated by Majorana bound states. 

Concerning the QME, the sweet spot regime of the I2KM behaves rather trivially when compared to the general case, reducing the problem to an effective single dot coupled to one ($\theta=0$) \cite{Lu2017} or two ($\theta=\pi/2$) leads \cite{Chatterjee2023}. On the other hand, the transport properties of the I2KM, or in general of an $N$-site Kitaev chain (for $N=3$, see Ref.~\cite{Bordin_2024}), 
coupled to two leads in different geometries represent an interesting topic per se.

\section{V. Non-uniqueness of the distance functions}\label{sec5}

The distance function in parameter space (M1) and the state distance function (M2) are not unique. However, similarly to the classical ME \cite{Lu2017}, our definition of the QME is indifferent to the specific choice of distance functions as long as they satisfy certain consistency conditions.

Concerning the distance in parameter space, for the classical (thermal) direct ME, the concept of distance follows by demanding that $T_{\rm eq} < T_c < T_h$  (or $T_c < T_h < T_{\rm eq}$ for the inverse ME). This ME is not affected by a redefinition of the distance function as long as the order remains preserved, i.e., for monotonic functions of temperature. 
In general, this corresponds to a local stretch of the temperature axis. Note that such a stretch would however affect the case $T_c < T_{\rm eq} < T_h$. Indeed, $|T_c-T_{\rm eq}|$ can be greater or smaller than $|T_f-T_{\rm eq}|$ depending on the used (monotonic) distance function. However, this case does not fall within the standard definition of the ME and has to be treated separately, see, e.g., Ref.~\cite{Tejero2024}.  Transformations not preserving the relative order should thus be excluded (and have been excluded throughout this work).  
A similar consideration can be given for non-thermal (or isothermal) MEs, where the quench in temperature is replaced by a quench in the strength of an external driving field \cite{Degunther2022} or in other parameters (at fixed temperature) \cite{Rylands2023}. 

If several parameters are quenched at the same time, the distance function in parameter space must recover the properties discussed above for a single quenched parameter.  The Euclidean distance measure (M1), or any distance measure obtained by stretching each parameter axis by means of a monotonic function, provides a valid and physically equivalent choice satisfying this condition. Indeed, all such measures will reduce to the condition $p_{\rm eq} < p_c < p_h$ (for the direct Mpemba effect, and similarly for the inverse one) if only one parameter $p$ is changed. Such stretches do not switch the order of "close" and "far" with respect of the equilibrium point as long as $p_{\rm eq}$ is lower (or higher) than both $p_c$ and $p_f$. Similarly to the classical ME, changing the distance function could therefore only affect situations with $p_c < p_{\rm eq} < p_f$ (or $p_c > p_{\rm eq} > p_f$). In analogy to the classical thermal ME, we exclude such cases here.

Also the distance function in quantum state space must be compatible with reasonable physical requests.  Following the arguments given in Ref.~\cite{Lu2017}, in order to avoid false reports of the QME, it must be a monotonically non-increasing, continuous, and convex function of time.  The identification of the QME  (including whether it is of type-I or type-II) is then indifferent to the specific choice of the distance function as long as these conditions are satisfied.
The trace distance (M2) satisfies all these conditions under  Markovian dynamics \cite{Nielsen2000,Wang2009}. The same properties are also satisfied by the relative entropy \cite{ruskai2022, hermes2017} and by the Bures distance \cite{spehner2013, petz1996}, while the Hilbert-Schmidt distance is convex \cite{tamir2015} but not always monotonic \cite{Wang2009}.
  
If the system dynamics is non-Markovian,  convexity of the trace distance (and of the other distance functions discussed above) is not guaranteed due the presence of information backflow from the environment to the system \cite{breuer2009, luo2023}.  Nonetheless, with suitable modifications, our approach may be adapted to the non-Markovian case. This requires to either find a monotonic and convex distance function replacing the trace distance, or to redefine the type-II QME (while still using the trace distance) such that the time $t^*$ (introduced in the main text) is the last, and not the first, crossing point between the two trajectories corresponding to the far and close initial conditions. However, we can not exclude at present that the identification of the QME will then depend on the choice of the distance function, depending on the master equation under consideration. A careful study of the QME for non-Markovian systems is an interesting topic for future research.

\bibliography{biblio}
\end{document}